\documentclass[10pt,twocolumn,aps,prl,superscriptaddress,nofootinbib,floatfix]{revtex4-2}
\usepackage{amsmath}
\usepackage{amssymb}
\usepackage{graphicx}
\usepackage{xcolor}
\usepackage{subfigure}
\usepackage{booktabs}
\usepackage{siunitx}
\usepackage{lineno}
\usepackage{hyperref}

\newcommand{\AFFicrr}{\affiliation{Kamioka Observatory, Institute for Cosmic Ray Research, University of Tokyo, Kamioka, Gifu 506-1205, Japan}}
\newcommand{\AFFkashiwa}{\affiliation{Research Center for Cosmic Neutrinos, Institute for Cosmic Ray Research, University of Tokyo, Kashiwa, Chiba 277-8582, Japan}}
\newcommand{\AFFipmu}{\affiliation{Kavli Institute for the Physics and
Mathematics of the Universe (WPI), The University of Tokyo Institutes for Advanced Study,
University of Tokyo, Kashiwa, Chiba 277-8583, Japan }}
\newcommand{\AFFmad}{\affiliation{Department of Theoretical Physics, University Autonoma Madrid, 28049 Madrid, Spain}}

\newcommand{\AFFbu}{\affiliation{Department of Physics, Boston University, Boston, MA~02215, USA}}
\newcommand{\AFFuci}{\affiliation{Department of Physics and Astronomy, University of California, Irvine, Irvine, CA~92697-4575, USA~}}
\newcommand{\AFFcsu}{\affiliation{Department of Physics, California State University, Dominguez Hills, Carson, CA~90747, USA}}
\newcommand{\AFFcnm}{\affiliation{Institute for Universe and Elementary Particles, Chonnam National University, Gwangju 61186, Korea}}
\newcommand{\AFFduke}{\affiliation{Department of Physics, Duke University, Durham NC 27708, USA}}
\newcommand{\AFFgifu}{\affiliation{Department of Physics, Gifu University, Gifu, Gifu 501-1193, Japan}}
\newcommand{\AFFgist}{\affiliation{GIST College, Gwangju Institute of Science and Technology, Gwangju 500-712, Korea}}
\newcommand{\AFFuh}{\affiliation{Department of Physics and Astronomy, University of Hawaii, Honolulu, HI 96822, USA}}
\newcommand{\AFFicl}{\affiliation{Department of Physics, Imperial College London , London, SW7 2AZ, United Kingdom }}
\newcommand{\AFFkek}{\affiliation{High Energy Accelerator Research Organization (KEK), Tsukuba, Ibaraki 305-0801, Japan }}
\newcommand{\AFFkobe}{\affiliation{Department of Physics, Kobe University, Kobe, Hyogo 657-8501, Japan}}
\newcommand{\AFFkyoto}{\affiliation{Department of Physics, Kyoto University, Kyoto, Kyoto 606-8502, Japan}}
\newcommand{\AFFliv}{\affiliation{Department of Physics, University of Liverpool, Liverpool, L69 7ZE, United Kingdom}}
\newcommand{\AFFmiyagi}{\affiliation{Department of Physics, Miyagi University of Education, Sendai, Miyagi 980-0845, Japan}}
\newcommand{\AFFnagoya}{\affiliation{Institute for Space-Earth Environmental Research, Nagoya University, Nagoya, Aichi 464-8602, Japan}}
\newcommand{\AFFkmi}{\affiliation{Kobayashi-Maskawa Institute for the Origin of Particles and the Universe, Nagoya University, Nagoya, Aichi 464-8602, Japan}}
\newcommand{\AFFpol}{\affiliation{National Centre For Nuclear Research, 02-093 Warsaw, Poland}}
\newcommand{\AFFsuny}{\affiliation{Department of Physics and Astronomy, State University of New York at Stony Brook, NY 11794-3800, USA}}
\newcommand{\AFFokayama}{\affiliation{Department of Physics, Okayama University, Okayama, Okayama 700-8530, Japan }}

\newcommand{\AFFox}{\affiliation{Department of Physics, Oxford University, Oxford, OX1 3PU, United Kingdom}}

\newcommand{\AFFseoul}{\affiliation{Department of Physics and Astronomy, Seoul National University, Seoul 151-742, Korea}}
\newcommand{\AFFsheff}{\affiliation{School of Mathematical and Physical Sciences, University of Sheffield, S3 7RH, Sheffield, United Kingdom}}
\newcommand{\AFFshizuokasc}{\affiliation{Department of Informatics in
Social Welfare, Shizuoka University of Welfare, Yaizu, Shizuoka, 425-8611, Japan}}
\newcommand{\AFFstfc}{\affiliation{STFC, Rutherford Appleton Laboratory, Harwell Oxford, and Daresbury Laboratory, Warrington, OX11 0QX, United Kingdom}}
\newcommand{\AFFskk}{\affiliation{Department of Physics, Sungkyunkwan University, Suwon 440-746, Korea}}
\newcommand{\AFFtodai}{\affiliation{Department of Physics, University of Tokyo, Bunkyo, Tokyo 113-0033, Japan }}
\newcommand{\AFFtit}{\affiliation{Department of Physics, Institute of Science Tokyo, Meguro, Tokyo 152-8551, Japan }}
\newcommand{\AFFtus}{\affiliation{Department of Physics and Astronomy, Faculty of Science and Technology, Tokyo University of Science, Noda, Chiba 278-8510, Japan }}
\newcommand{\AFFtriumf}{\affiliation{TRIUMF, 4004 Wesbrook Mall, Vancouver, BC, V6T2A3, Canada }}

\newcommand{\AFFtsinghua}{\affiliation{Department of Engineering Physics, Tsinghua University, Beijing, 100084, China}}
\newcommand{\AFFynu}{\affiliation{Department of Physics, Yokohama National University, Yokohama, Kanagawa, 240-8501, Japan}}
\newcommand{\AFFllr}{\affiliation{Ecole Polytechnique, IN2P3-CNRS, Laboratoire Leprince-Ringuet, F-91120 Palaiseau, France }}
\newcommand{\AFFbari}{\affiliation{ Dipartimento Interuniversitario di Fisica, INFN Sezione di Bari and Universit\`a e Politecnico di Bari, I-70125, Bari, Italy}}
\newcommand{\AFFnapoli}{\affiliation{Dipartimento di Fisica, INFN Sezione di Napoli and Universit\`a di Napoli, I-80126, Napoli, Italy}}
\newcommand{\AFFroma}{\affiliation{INFN Sezione di Roma and Universit\`a di Roma ``La Sapienza'', I-00185, Roma, Italy}}
\newcommand{\AFFpadova}{\affiliation{Dipartimento di Fisica, INFN Sezione di Padova and Universit\`a di Padova, I-35131, Padova, Italy}}
\newcommand{\AFFkeio}{\affiliation{Department of Physics, Keio University, Yokohama, Kanagawa, 223-8522, Japan}}
\newcommand{\AFFwinnipeg}{\affiliation{Department of Physics, University of Winnipeg, MB R3J 3L8, Canada }}
\newcommand{\AFFkcl}{\affiliation{Department of Physics, King's College London, London, WC2R 2LS, UK }}
\newcommand{\AFFwarwick}{\affiliation{Department of Physics, University of Warwick, Coventry, CV4 7AL, UK }}
\newcommand{\AFFral}{\affiliation{Rutherford Appleton Laboratory, Harwell, Oxford, OX11 0QX, UK }}
\newcommand{\AFFwu}{\affiliation{Faculty of Physics, University of Warsaw, Warsaw, 02-093, Poland }}
\newcommand{\AFFbcit}{\affiliation{Department of Physics, British Columbia Institute of Technology, Burnaby, BC, V5G 3H2, Canada }}
\newcommand{\AFFtohoku}{\affiliation{Department of Physics, Faculty of Science, Tohoku University, Sendai, Miyagi, 980-8578, Japan }}
\newcommand{\AFFicise}{\affiliation{Institute For Interdisciplinary Research in Science and Education, ICISE, Quy Nhon, 55121, Vietnam }}
\newcommand{\AFFilance}{\affiliation{ILANCE, CNRS - University of Tokyo International Research Laboratory, Kashiwa, Chiba 277-8582, Japan}}
\newcommand{\AFFibs}{\affiliation{Center for Underground Physics, Institute for Basic Science (IBS), Daejeon, 34126, Korea}}
\newcommand{\AFFglasgow}{\affiliation{School of Physics and Astronomy, University of Glasgow, Glasgow, Scotland, G12 8QQ, United Kingdom}}
\newcommand{\AFFoecu}{\affiliation{Media Communication Center, Osaka Electro-Communication University, Neyagawa, Osaka, 572-8530, Japan}}
\newcommand{\AFFminn}{\affiliation{School of Physics and Astronomy, University of Minnesota, Minneapolis, MN  55455, USA}}
\newcommand{\AFFsilesia}{\affiliation{August Che\l{}kowski Institute of Physics, University of Silesia in Katowice, 75 Pu\l{}ku Piechoty 1, 41-500 Chorz\'{o}w, Poland}}
\newcommand{\AFFtoyama}{\affiliation{Faculty of Science, University of Toyama, Toyama City, Toyama 930-8555, Japan}}
\newcommand{\AFFbmcc}{\affiliation{Science Department, Borough of Manhattan Community College / City University of New York, New York, New York, 1007, USA.}}
\newcommand{\AFFnumazu}{\affiliation{National Institute of Technology, Numazu College, Numazu, Shizuoka  410-8501, Japan}}
\newcommand{\AFFniihama}{\affiliation{National Institute of Technology, Niihama College, Niihama, Ehime  792-8580, Japan}}
\newcommand{\AFFucas}{\affiliation{School of Physical Sciences, University of Chinese Academy of Sciences, Beijing 101408, China}}

\begin{document}

\title{First observation of electron antineutrinos from nuclear reactors at Super-Kamiokande}


\author{K.~Abe}
\AFFicrr
\AFFipmu
\author{Y.~Asaoka}
\AFFicrr
\AFFipmu
\author{M.~Harada}
\AFFicrr
\author{Y.~Hayato}
\AFFicrr
\AFFipmu
\author{K.~Hiraide}
\AFFicrr
\AFFipmu
\author{T.~H.~Hung}
\AFFicrr
\author{K.~Ieki}
\author{M.~Ikeda}
\AFFicrr
\AFFipmu
\author{J.~Kameda}
\AFFicrr
\AFFipmu
\author{Y.~Kataoka}
\AFFicrr
\AFFipmu
% ...


\collaboration{The Super-Kamiokande Collaboration}
\noaffiliation

\begin{abstract}

In 2020, Super-Kamiokande entered the SK-Gd phase, improving its sensitivity to low-energy electron antineutrinos through the detection of neutrons captured on {Gadolinium} dissolved in the detector water. An analysis of reactor electron antineutrino data has been carried out using data from SK-Gd, with a~total exposure of 411.52 live days. The observed prompt energy spectra and monthly event rates are consistent with the oscillated reactor antineutrino prediction after accounting for accidental, geoneutrino, and spallation backgrounds. The no-reactor hypothesis is disfavored with a~reactor-antineutrino signal significance of approximately $6\sigma$. The oscillation parameters are measured to be $\sin^2\theta_{12} = 
0.500\pm0.155$ 
and $\Delta m^2_{21} = (9.08_{-0.54}^{+0.55}) \times 10^{-5}~\mathrm{eV}^2$. A~comparison with Super-Kamiokande solar neutrino measurements shows a statistical compatibility between the two datasets at the $1.9\sigma$ level. A~combined solar and reactor fit in Super-Kamiokande yields best-fit values of $\sin^2\theta_{12} = 0.332_{-0.026}^{+0.028}$ and $\Delta m^2_{21} = (8.94_{-0.47}^{+0.46}) \times 10^{-5}~\mathrm{eV}^2$.
\end{abstract}

\maketitle

\textit{Introduction ---}
Reactor electron antineutrinos are produced during the beta decay of fission products in nuclear reactors,
\begin{equation}
    (A,Z) \rightarrow (A,Z+1) + e^- + \bar{\nu}_e,
\end{equation}
in which a~neutron inside the nucleus undergoes beta decay, producing a proton, an electron $e^-$, and an electron antineutrino $\bar{\nu}_e$.
{Reactors were the first source of neutrinos detected in the 1956 Cowan--Reines experiment~\citep{Cowan:1956rrn}, which confirmed their role in beta decays.}
Since then, many studies of reactor neutrinos have improved our knowledge of neutrino properties, including their oscillation parameters and {the search for additional interactions} such as a~neutrino magnetic moment~\citep{VogelEngel1989,Beda2012}.

Since 2020, Super-Kamiokande (Super-K) has operated in the SK-Gd phase, in which Gadolinium dissolved in the detector water enhances neutron detection following inverse beta decay (IBD) interactions, improving the sensitivity to low-energy electron antineutrinos~\citep{Super-Kamiokande:2021the}.
Several nuclear power plants are located about 200~km of Super-K, with the observed reactor antineutrinos primarily produced by Japanese nuclear power plants and a~smaller contribution from reactors in Korea. Some of the closest active nuclear reactors to Super-K are located at Wakasa Bay: Mihama unit 3 (146~km), Ohi units 3 and 4 (179~km), and Takahama units 1, 2, 3, and 4 (191~km). During the first years of SK-Gd (2020--2023), many of these reactors were either active or in the process of restarting. 
KamLAND, located in the same underground complex as Super-Kamiokande, previously studied reactor antineutrinos in a similar baseline range from Japanese nuclear reactors~\citep{KamLAND2003,KamLAND2005}.

The reactor and geoneutrino flux predictions used in this work are obtained with the framework described in~\citep{geoneutrinospaperref,geoneutrinos_accessed}. The calculation uses reactor activity information from the IAEA Power Reactor Information System (PRIS)~\citep{IAEA_PRIS} to account for the time-dependent contribution of individual reactors.

In this work, we present an analysis of reactor electron antineutrino data from SK-Gd. The enhanced neutron detection capability provided by {Gadolinium} loading enables precise measurements of the prompt energy spectrum and time-dependent event rate, leading to constraints on the solar oscillation parameters, which are compared with solar neutrino measurements to test CPT invariance, and combined with them to further improve the accuracy of the constraints.

\textit{Super-Kamiokande Detector ---}
The Super-K experiment is a~50~kiloton (kt) water Cherenkov detector located 1000~m underground in the Kamioka mine in Japan~\citep{Super-Kamiokande:2002weg}. {The detector consists of an inner detector containing 32~kt of water viewed by 11,129 inward-facing 20-inch photomultiplier tubes (PMTs), surrounded by an outer detector.} The usual fiducial volume used in SK analyses is 22.5~kt. Charged particles traversing the detector produce Cherenkov light, allowing reconstruction of the event vertex, direction, and energy.

The SK-Gd phase was implemented in two stages. In July 2020, 13~tons of Gadolinium sulfate octahydrate ($\mathrm{Gd_2(SO_4)_3 \cdot 8H_2O}$) were dissolved into the water of the detector, achieving a~{Gadolinium} concentration of 0.011\% by mass~\citep{Super-Kamiokande:2021the}. This marked the start of the SK-VI phase. In June 2022, an additional 27~tons of the compound were added, increasing the concentration to 0.033\% by mass~\citep{ABE2024169480} and initiating the SK-VII phase.

An IBD interaction produces a~prompt positron, detected via Cherenkov radiation, while an accompanying neutron undergoes thermalization in water before being captured by nuclei. The resulting nuclear de-excitation emits $\gamma$~rays, which are observed mainly through Compton-scattered electrons producing Cherenkov light.
The energy of the de-excitation $\gamma$~rays depends on the capture nucleus. A~neutron capture on hydrogen emits a~single 2.2~MeV $\gamma$~ray, while captures on Gd emit easily detectable $\gamma$~ray cascades of about 8~MeV~\citep{PhysRevLett.93.171101}. The fraction of neutron captures on {Gadolinium} is expected to be approximately 50\% in SK-VI and increases to about 75\% in SK-VII, reflecting the higher Gd loading in this phase. The presence of Gd also shortens the neutron capture time, expected to be $\sim120~\si{\micro\second}$ for SK-VI and $\sim60~\si{\micro\second}$ for SK-VII~\citep{ABE2024169480}.

In addition to {Gadolinium} loading, the ability to efficiently detect low-energy neutrinos in Super-K also relies on advances in the trigger system. The Wideband Intelligent Trigger (WIT)~\citep{CARMINATI2015666} provides efficient selection of low-energy events with thresholds down to a~few MeV. This capability enables searches for solar neutrinos and {pre-supernova (preSN) neutrinos~\citep{Kato_2020} (see below)}, and is essential for detecting the reactor antineutrino signal studied in this work.

\textit{Analysis Strategy ---}
The search for reactor electron antineutrinos in Super-K builds on techniques developed for the preSN neutrino analysis~\citep{Machado_2022,combined_preSN,charlespaper}, {which targets low-energy electron antineutrinos emitted during the final silicon-burning stage of massive stars, hours before a~core-collapse supernova.} Both preSN and reactor $\bar{\nu}_e$ are expected in a~similar energy range and are detected via IBD.
The analysis uses low-energy events recorded by the WIT system. WIT receives the same detector data stream as the standard trigger and processes it in real time using a~dedicated computing farm. Detector data are analyzed in 23~ms blocks with a low-threshold trigger algorithm that requires approximately 11 hits above the dark-noise expectation within a 230~ns window, providing near-100\% efficiency for electrons above about 2.5~MeV. Candidate events passing this initial selection undergo vertex reconstruction and quality checks before being organized in time and made available for offline low-energy analyses.

Reactor antineutrino candidates are identified through the prompt--delayed coincidence signature of the IBD interaction. The prompt positron signal and delayed neutron-capture signal form a~characteristic coincidence signature, whose time interval and spatial separation, referred to as the coincidence time and coincidence distance, provide strong discrimination against background events.

{Signal events are simulated by modeling the detector response to both the prompt positron and delayed neutron-capture signals.} The detector response to neutron capture on Gadolinium is modeled by propagating $\gamma$~rays generated according to dedicated emission models, while positrons with energies between 0.8 and 10.0~MeV are simulated independently. 
For each simulated IBD event, the prompt and delayed signals are combined, and their reconstructed spatial and temporal separation is used in the analysis.


The expected reactor antineutrino signal, including oscillations, is obtained by weighting the simulated events using energy spectra from the global antineutrino framework described in~\citep{geoneutrinospaperref,geoneutrinos_accessed}. This framework provides reactor and geoneutrino interaction-rate and energy-spectrum predictions at arbitrary locations on Earth and is used here to evaluate the fluxes at Super-K. Reactor activity is determined using PRIS, which provides information on reactor specifications, performance, and operational history. The calculation uses monthly reactor load factors from PRIS to account for the time-dependent contribution of individual reactors. These load factors are proportional to the electrical power output, while the antineutrino emission is directly related to the reactor \emph{thermal} power, introducing an intrinsic uncertainty that is not completely captured by the available data.

Reactor antineutrino emission spectra are constructed following the methodology described in~\citep{GIUNTI2022137054}, with different models considered for the isotopic antineutrino spectra of ${}^{235}\mathrm{U}$, ${}^{238}\mathrm{U}$, ${}^{239}\mathrm{Pu}$, and ${}^{241}\mathrm{Pu}$. Two spectral model approaches are considered: the Huber-Kopeikin model, which combines the predictions of~\citep{Kopeikin2021,Huber2011} in the 2--8~MeV range with~\citep{Estienne} outside this region, and the Huber-Mueller model, which uses~\citep{Mueller} for ${}^{238}\mathrm{U}$ and~\citep{Huber2011} for ${}^{235}\mathrm{U}$, ${}^{239}\mathrm{Pu}$, and ${}^{241}\mathrm{Pu}$ in the 2--8~MeV range, again combined with~\citep{Estienne} outside this region. The differences between these models are treated as a~source of systematic uncertainty. The event rates are calculated using the IBD cross section from~\citep{StrumiaVissani2003}.

IBD candidates are searched for in the reconstructed data sample by identifying prompt--delayed event pairs. The reconstructed prompt energy is defined as the total visible energy of the prompt positron event and is used to construct the energy spectrum employed in the oscillation analysis.
For each prompt candidate, a delayed candidate is searched in a time window from $-17~\si{\micro\second}$ to $290~\si{\micro\second}$ relative to the prompt event. Candidate pairs are characterized by the coincidence time ($dT$) and coincidence distance ($dR$), together with reconstructed properties of the delayed neutron-capture signal, including its reconstruction quality and the number of PMT hits within an 18~ns timing window ($n18$). These variables are combined using a Boosted Decision Tree (BDT) classifier trained to separate simulated IBD events from background-dominated data samples. The final event selection is based on the BDT response, together with the requirements $dR<200$~cm and $dT<80~\si{\micro\second}$. The selection criteria are optimized to maximize the sensitivity to reactor antineutrino signals.




Several background sources contribute in the energy region relevant for reactor antineutrino detection. These backgrounds include genuine electron antineutrinos, which constitute an irreducible background, and events that mimic the IBD signature.

Geoneutrinos are antineutrinos produced by natural radioactivity within the Earth and constitute an irreducible background to reactor antineutrino detection. The geoneutrino flux in the energy region relevant to this analysis is expected to be small. The geoneutrino electron antineutrino flux in Super-K was evaluated using~\citep{geoneutrinospaperref,geoneutrinos_accessed}, with main contributions arising from the mantle fluxes associated with the ${}^{238}$U and ${}^{232}$Th decay chains.

Another background arises from accidental coincidences that occur when otherwise uncorrelated events are sufficiently close in time and space to mimic the prompt--delayed signature of an IBD interaction. These coincidences represent an intrinsic detector background and cannot be completely removed by the event selection.

Cosmic-ray spallation products are produced when high-energy muons interact with nuclei in the detector, producing unstable radioactive isotopes that decay shortly after the muon's passage. The resulting events can form clusters in time and can therefore mimic IBD candidates. These backgrounds are efficiently reduced through a combination of the BDT-based event selection and the rejection of events with a preceding muon within 0.3~ms.

{The background estimation relies on periods of low reactor activity in Japan. The period from November to December 2020 was characterized by minimal reactor operation, with the main reactors in the Wakasa Bay region either inactive or undergoing shutdown. As a~result, the expected reactor antineutrino flux was strongly suppressed, and the data collected during this period provide a~sample dominated by background events.}

The observed prompt-energy spectrum during this low-reactor period is used as the basis for the accidental background estimate. Small residual contributions from reactor antineutrinos and geoneutrinos are subtracted, and events identified as spallation-like are removed. To propagate the statistical uncertainty of the low-reactor data and the uncertainties associated with the residual reactor and geoneutrino subtraction, a~set of toy Monte Carlo samples is generated by statistically fluctuating these contributions. The resulting spectra are scaled to the full analysis live time, with the mean in each reconstructed prompt-energy bin used as the accidental background prediction and the spread of the toy spectra used to estimate its uncertainty, which is dominated by the statistical uncertainty of the low-reactor data sample.

Systematic uncertainties affecting the predicted spectra were evaluated and propagated to the oscillation analysis. These include uncertainties associated with the reactor antineutrino flux and spectrum models, the geoneutrino prediction, the IBD cross section, and the modeling of neutron-capture $\gamma$ rays. Reactor-model uncertainties were estimated by comparing predictions obtained using the Huber-Kopeikin and Huber-Mueller models~\citep{GIUNTI2022137054,Kopeikin2021,Huber2011,Mueller}. The IBD cross-section uncertainty was propagated from~\citep{StrumiaVissani2003}. The uncertainty associated with neutron-capture $\gamma$ rays was evaluated from the differences between predictions obtained using the GLG4sim~\citep{chooz_ref} and ANNRI~\citep{anri_ref_sim} $\gamma$~ray emission models, following the procedure described in~\citep{Machado_2022}.

Detector-related uncertainties include trigger efficiency, live time, and event-selection efficiency. The trigger-efficiency uncertainty was evaluated using calibration data from a~Ni-Cf source deployed at different positions in the detector and was found to be consistent with estimates obtained from dedicated Monte Carlo studies.

\textit{Results ---}
A~full analysis of both SK-VI and SK-VII data was carried out, comparing the observed data to estimations based on background modeling and reactor Monte Carlo simulations. The SK-VI sample covers the period from February 2021 to May 2022, with 229.05 days of effective {live time} used in this analysis. The SK-VII sample spans September 2022 to September 2023, contributing 182.47 days of effective {live time}, for a total exposure of 411.52 days. SK-VI data collected between November and December 2020 were used to estimate the accidental background contribution, as reactor activity was particularly low during this period.
In both phases, only runs taken under stable detector conditions were included, and periods flagged as bad runs were excluded from the analysis.

\begin{figure*}[tbp]
\centering
\subfigure[SK-VI]{%
\includegraphics[width=0.48\textwidth]{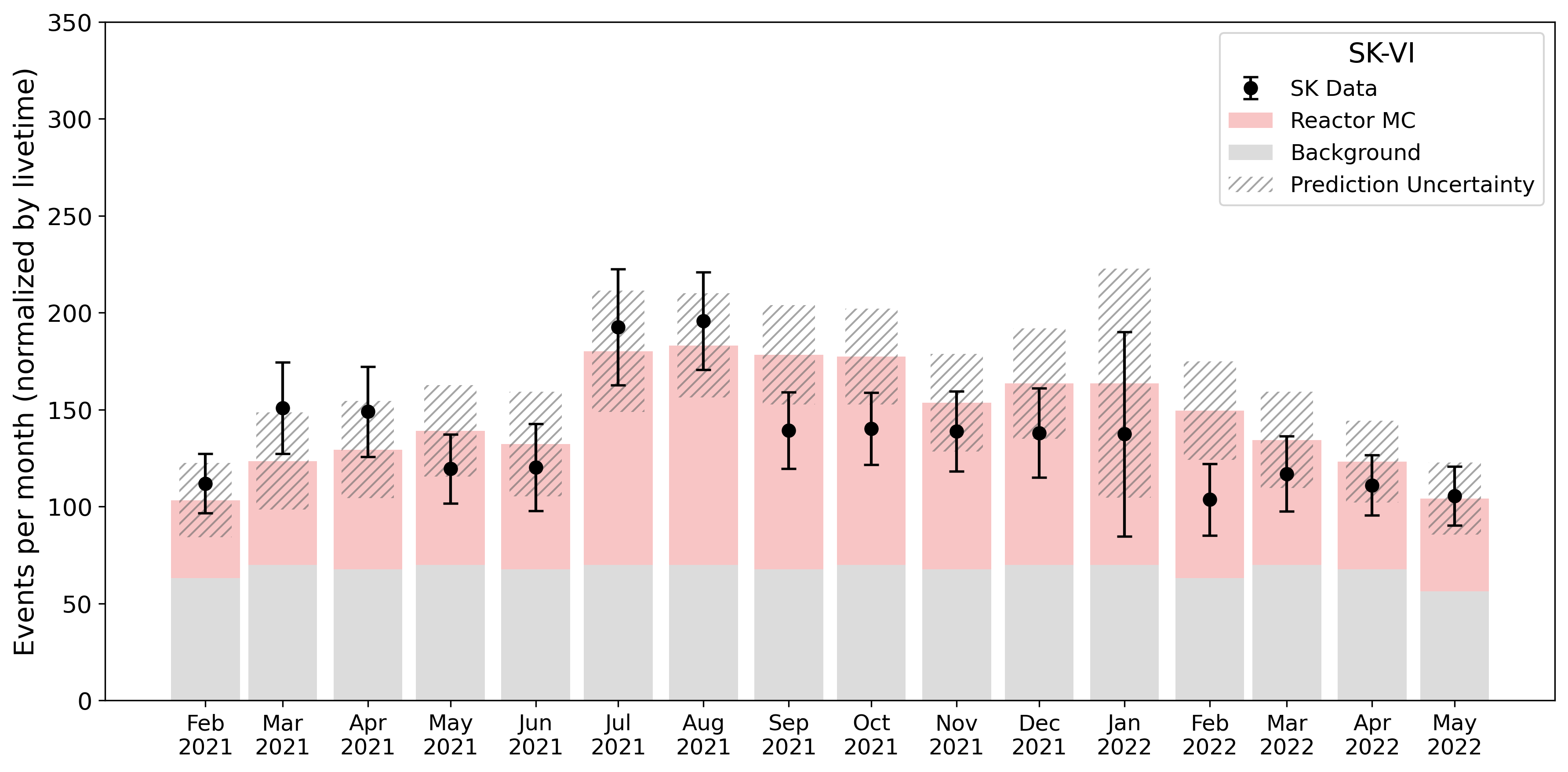}
}
\hfill
\subfigure[SK-VII]{%
\includegraphics[width=0.48\textwidth]{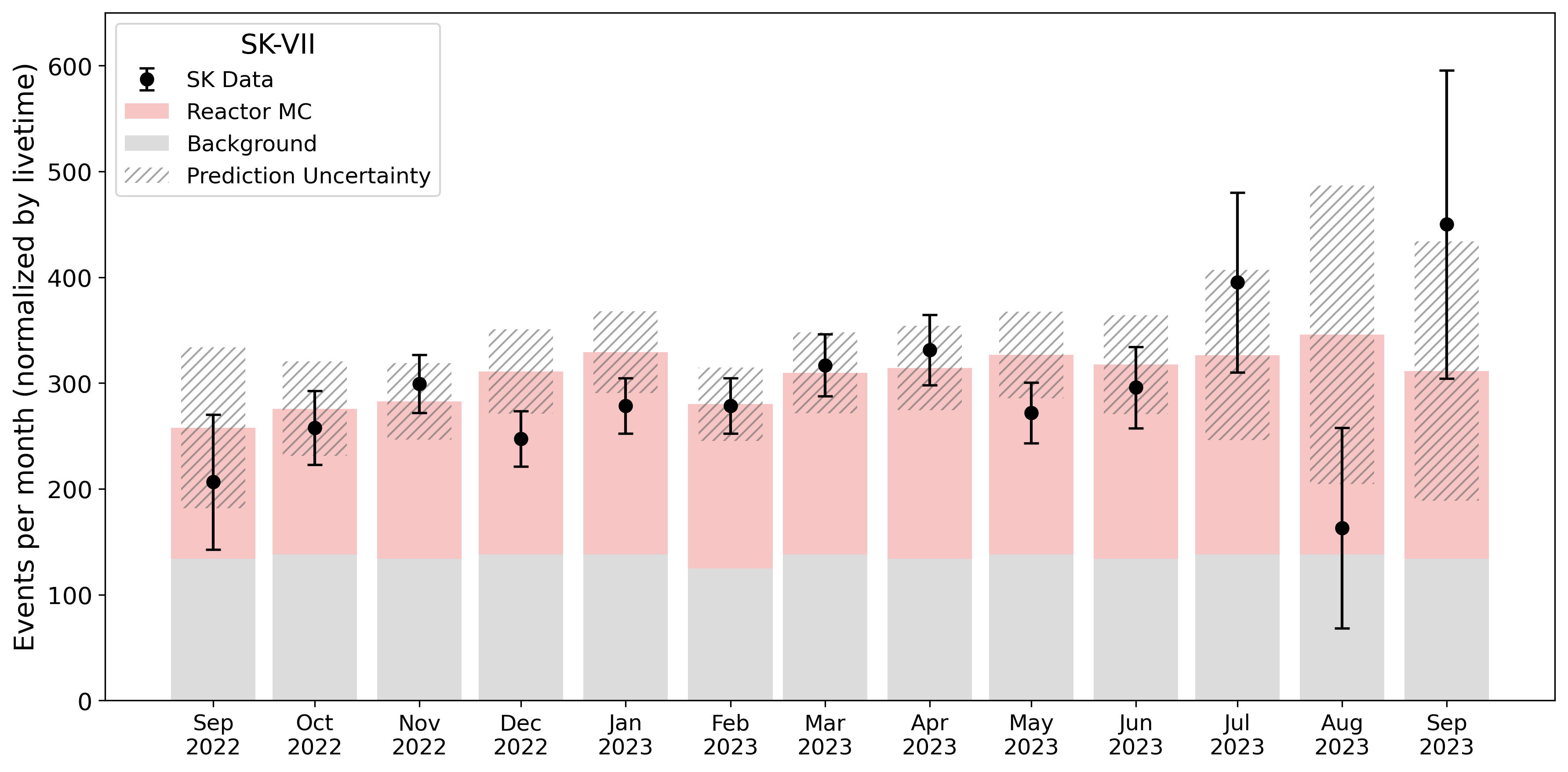}
}
\caption{Monthly observed and predicted event rates for SK-VI and SK-VII, normalized to a~full calendar month using the selected {live time} and excluding known bad-quality periods. Stacked bars show reactor and background predictions; black markers show data with total uncertainties; hatched bands show prediction uncertainties.}
\label{fig:monthlyrates_skvi_skvii}
\end{figure*}

The reactor antineutrino predictions were obtained using a~local copy of~\citep{geoneutrinospaperref, geoneutrinos_accessed}. Figure~\ref{fig:monthlyrates_skvi_skvii} shows the monthly event rates for both SK-VI and SK-VII, compared to the expected reactor contribution calculated using the best-fit oscillation parameters and the expected background from accidental coincidences and geoneutrinos. A~month-by-month comparison yields a~mean p-value of 0.41, indicating that the monthly observed rates are consistent with the prediction within uncertainties, {while the corresponding mean p-values are 0.067 for the background-only hypothesis and 0.0089 for the background plus unoscillated reactor-antineutrino hypothesis.} 
For January 2022, September 2022, and July, August, and September 2023, the {uncertainties} are noticeably larger. This is due to reduced {live time} in these months, mainly caused by detector maintenance, periods of convection, or unstable data-taking conditions. Since the event rates are normalized to a~full calendar month, the statistical uncertainties are correspondingly scaled, resulting in larger {uncertainties} compared to months with higher {live time}.

\begin{figure}[tbp]
\centering
\includegraphics[width=0.48\textwidth]{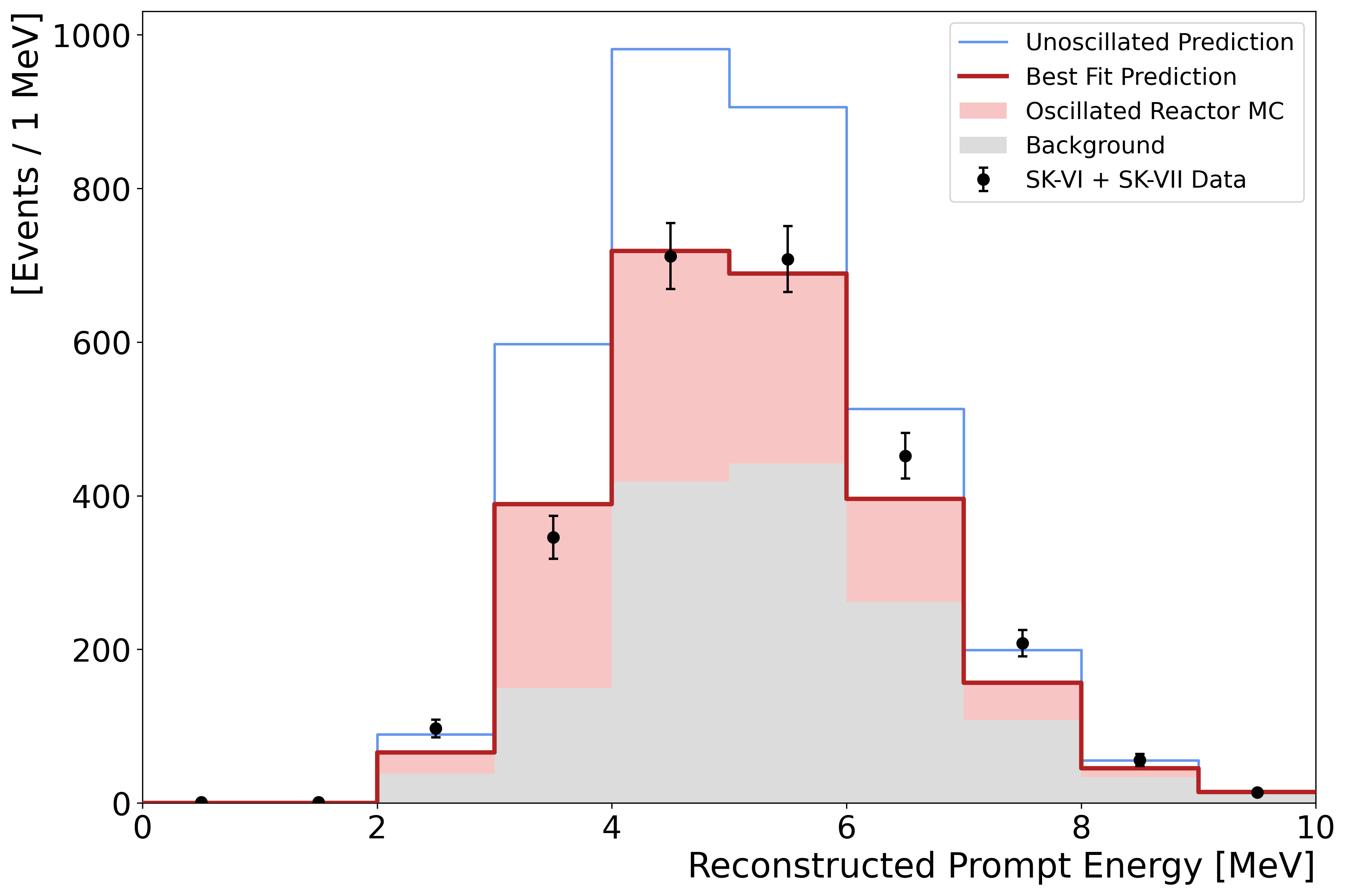}
\caption{Combined reconstructed prompt energy spectrum of observed data compared to model predictions for SK-VI and SK-VII. Black markers show the observed spectrum. The solid blue line shows the unoscillated prediction, while the stacked bars show the background and oscillated reactor contributions. The solid dark-red line shows the best-fit prediction, with $\sin^2\theta_{12}=0.500$ and $\Delta m^2_{21}=9.08\times10^{-5}~\mathrm{eV}^2$.}
\label{fig:spectrum_combined}
\end{figure}

Figure~\ref{fig:spectrum_combined} shows the prompt energy spectrum for the combined SK-VI and SK-VII periods compared against the estimated background and reactor signal. The prediction shown corresponds to the best-fit oscillation parameters.

\begin{figure}[tbp]
\centering
\includegraphics[width=0.45\textwidth]{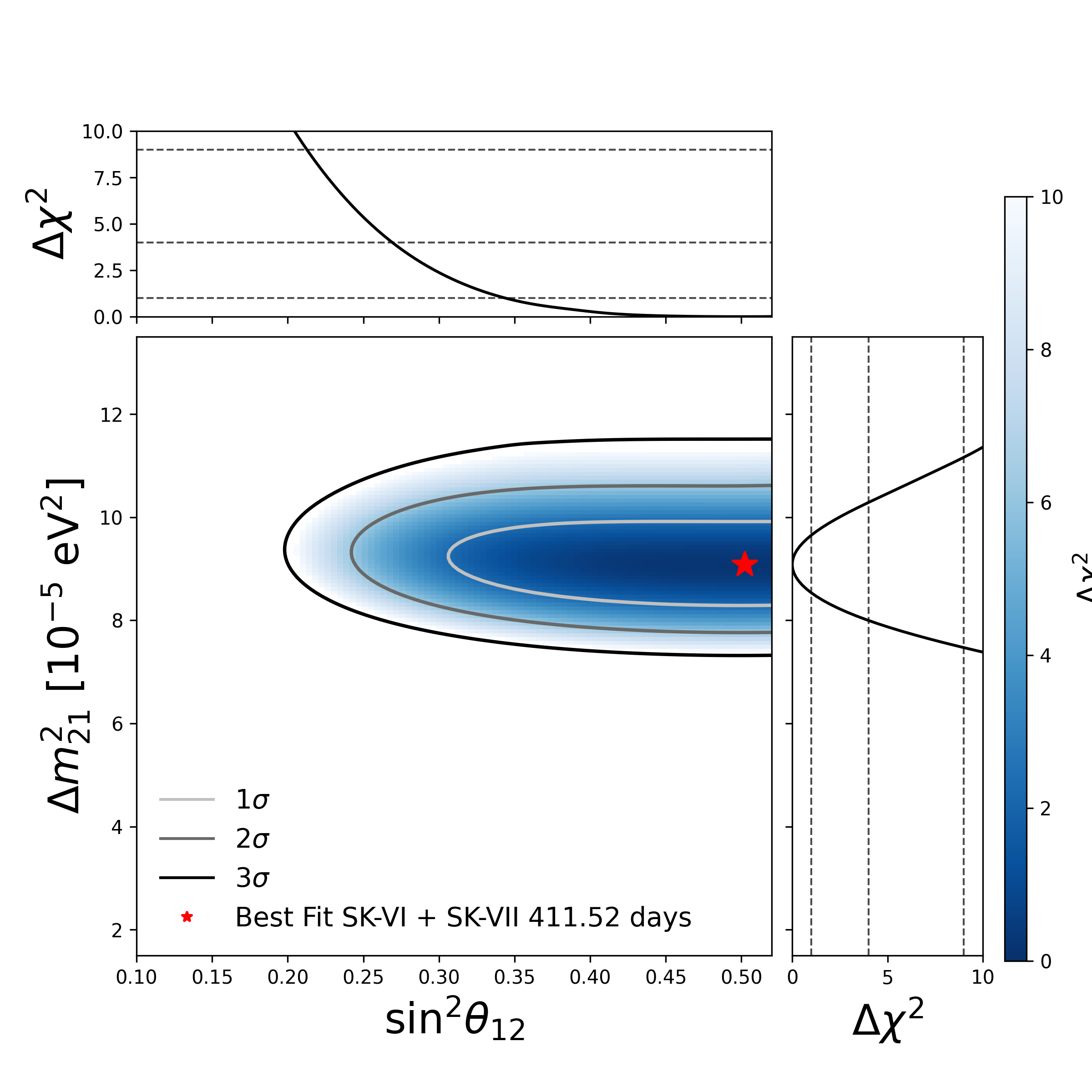}
\caption{Combined $\Delta\chi^2$ surface for SK-VI and SK-VII, showing the best-fit point (red star) and the 1$\sigma$, 2$\sigma$, and 3$\sigma$ confidence regions in the $(\sin^2\theta_{12},\Delta m^2_{21})$ parameter space.}
\label{fig:chisquared_gridSK}
\end{figure}

{To determine the best-fit oscillation parameters, the observed prompt-energy spectrum was compared with predictions over the ($\sin^2\theta_{12}$, $\Delta m^2_{21}$) parameter space by minimizing a~$\chi^2$ function that includes both statistical and systematic uncertainties.} The parameter $\sin^2\theta_{13}$ was fixed to the {NuFit v6.0~\citep{NUFIT60}} best-fit value of 0.02203. Figure~\ref{fig:chisquared_gridSK} shows the resulting combined $\chi^2$ surface obtained from the SK-VI and SK-VII datasets. Confidence contours corresponding to 1$\sigma$, 2$\sigma$, and 3$\sigma$ ($\Delta\chi^2=2.30$, 6.18, and 11.8, respectively) are shown together with the global minimum (red star). The uncertainties on the oscillation parameters were obtained from one-dimensional profile likelihoods. This procedure yields $\sin^2\theta_{12} =
0.500\pm0.155$ 
and $\Delta m^2_{21} = (9.08_{-0.54}^{+0.55}) \times 10^{-5}~\mathrm{eV}^2$.

The combined SK-VI and SK-VII datasets are well described by the oscillated reactor-antineutrino hypothesis, yielding a~goodness-of-fit value of $\chi^2 = 16.7$ for 16 degrees of freedom, corresponding to a~p-value of 0.40. The goodness-of-fit is evaluated using eight reconstructed prompt-energy bins from each SK-Gd phase. The no-reactor neutrino hypothesis is strongly disfavored, with $\Delta\chi^2=34.7$ relative to the oscillated reactor-antineutrino hypothesis, corresponding to a~reactor-antineutrino signal significance of approximately $6\sigma$. These results establish the observation of reactor antineutrinos in SK-Gd and allow the extraction of oscillation parameters for direct comparison with solar-neutrino measurements.

\begin{figure}[tbp]
\centering
\includegraphics[width=0.45\textwidth]{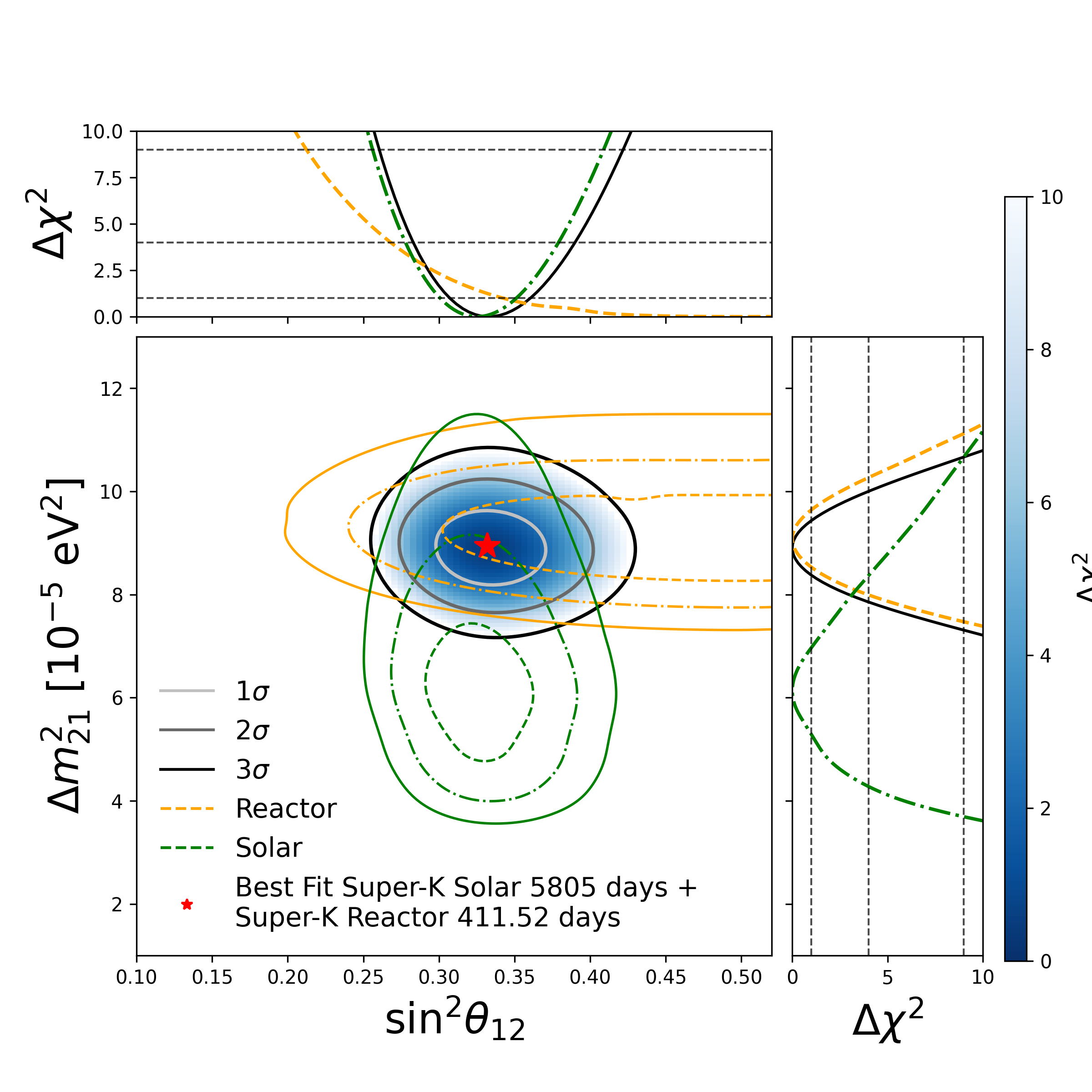}
\caption{Combined $\Delta\chi^2$ surface for the Super-K reactor and solar analyses, showing the best-fit point (red star) and the 1$\sigma$, 2$\sigma$, and 3$\sigma$ confidence regions in the $(\sin^2\theta_{12},\Delta m^2_{21})$ parameter space.}
\label{fig:chisquared_gridSK_withSolar}
\end{figure}

The oscillation parameters extracted from solar neutrinos and reactor antineutrinos within Super-K were also compared. The Super-K solar analysis~\citep{PhysRevD.109.092001} reports $\sin^2\theta_{12}^{(\nu)} = 0.324^{+0.027}_{-0.023}$ and $\Delta m^2_{21}{}^{(\nu)} = (6.10^{+1.26}_{-0.86}) \times 10^{-5},\mathrm{eV}^2$, while the reactor-only fit from this work gives $\sin^2\theta_{12}^{(\bar{\nu})} = 
0.500\pm0.155$ 
and $\Delta m^2_{21}{}^{(\bar{\nu})} = (9.08^{+0.55}_{-0.54}) \times 10^{-5},\mathrm{eV}^2$. 
The differences between the neutrino and antineutrino determinations correspond to approximately \(1.1\sigma\) for \(\sin^2\theta_{12}\) and \(2.2\sigma\) for \(\Delta m^2_{21}\). {These differences do not constitute statistically significant evidence for a difference in oscillation parameters (and therefore CPT violation).}

{A more rigorous assessment of the consistency between the solar and reactor measurements was performed using a parameter goodness-of-fit (PG) test, which evaluates the global compatibility of the two datasets within the oscillation framework. The resulting value, $\Delta\chi^2_{\mathrm{PG}} = 7.25$ for 2 degrees of freedom, corresponds to a p-value of 0.027, equivalent to a compatibility of the two datasets at the $1.9\sigma$ level. This indicates that the reactor and solar  measurements are still statistically compatible.}

{A~joint Super-K solar and reactor analysis was performed by combining the results from the Super-K solar neutrino analysis reported in~\citep{PhysRevD.109.092001}, {combining the reactor and solar $\chi^2$ surfaces.} Figure~\ref{fig:chisquared_gridSK_withSolar} shows the resulting confidence contours and best-fit point, yielding $\sin^2\theta_{12} = 0.332_{-0.026}^{+0.028}$ and $\Delta m^2_{21} = (8.94_{-0.47}^{+0.46}) \times 10^{-5}~\mathrm{eV}^2$. }

\textit{Conclusion ---}
An analysis of low-energy reactor antineutrino data in Super-Kamiokande has been carried out using both SK-VI and SK-VII phases, corresponding to a~total exposure of 411.52 live days. The observed prompt energy spectra and event rates were found to be consistent with the oscillated reactor antineutrino prediction after accounting for detector backgrounds.

The combined SK-VI and SK-VII datasets yield a~best fit consistent with neutrino oscillations, while both the unoscillated and no-reactor neutrino hypotheses are strongly disfavored. The extracted oscillation parameters are consistent with global fits and recent reactor-antineutrino measurements~\citep{NUFIT60,JUNO2026,SNOplus2026} and are determined with well-defined confidence intervals.

When combined with the Super-K solar neutrino analysis, the oscillation contours become further constrained, indicating the complementary sensitivities of solar and reactor measurements. {A~direct comparison (motivated by CPT invariance)} between solar neutrinos and reactor antineutrinos shows no statistically significant discrepancy.

These results establish the first observation of reactor electron antineutrinos in Super-Kamiokande and demonstrate the potential of {Gadolinium}-loaded water Cherenkov detectors for future low-energy antineutrino measurements.

\textit{Acknowledgments ---}
We gratefully acknowledge the cooperation of the Kamioka Mining and Smelting Company. The Super-Kamiokande experiment has been built and operated from funding by the Japanese Ministry of Education, Culture, Sports, Science and Technology; the U.S. Department of Energy; and the U.S. National Science Foundation. Some of us have been supported by funds from the National Research Foundation of Korea (NRF-2009-0083526, NRF-2022R1A5A1030700, NRF-2022R1A3B1078756, RS-2025-00514948, RS-2025-23525600, RS-2026-25489835) funded by the Ministry of Science, Information and Communication Technology (ICT); the Institute for Basic Science (IBS-R016-D1); and the Ministry of Education (2018R1D1A1B07049158, 2021R1I1A1A01042256, RS-2024-00442775); the Japan Society for the Promotion of Science; the National Natural Science Foundation of China (Grants No. 12375100 and 12521007); the Spanish Ministry of Science, Universities and Innovation (grant PID2024-157541NB-I00); the Natural Sciences and Engineering Research Council (NSERC) of Canada; the SciNet and the Digital Research Alliance of Canada; the National Science Centre (UMO-2022/46/E/ST2/00336) and the Ministry of  Science and Higher Education (2023/WK/04), Poland; the Science and Technology Facilities Council (STFC) and Grid for Particle Physics (GridPP), UK; the European Union's Horizon Europe Research and Innovation Programme HORIZON-MSCA-SE-2023 JENNIFER3 project, grant agreement no.101183137, H2020-MSCA-RISE-2019 SK2HK grant agreement no. 872549; and European Union's Next Generation EU/PRTR  grant CA3/RSUE2021-00559; the National Institute for Nuclear Physics (INFN), Italy.

\bibliographystyle{apsrev4-2}
\bibliography{references}

\begin{thebibliography}{29}%
\makeatletter
\providecommand \@ifxundefined [1]{%
 \@ifx{#1\undefined}
}%
\providecommand \@ifnum [1]{%
 \ifnum #1\expandafter \@firstoftwo
 \else \expandafter \@secondoftwo
 \fi
}%
\providecommand \@ifx [1]{%
 \ifx #1\expandafter \@firstoftwo
 \else \expandafter \@secondoftwo
 \fi
}%
\providecommand \natexlab [1]{#1}%
\providecommand \enquote  [1]{``#1''}%
\providecommand \bibnamefont  [1]{#1}%
\providecommand \bibfnamefont [1]{#1}%
\providecommand \citenamefont [1]{#1}%
\providecommand \href@noop [0]{\@secondoftwo}%
\providecommand \href [0]{\begingroup \@sanitize@url \@href}%
\providecommand \@href[1]{\@@startlink{#1}\@@href}%
\providecommand \@@href[1]{\endgroup#1\@@endlink}%
\providecommand \@sanitize@url [0]{\catcode `\\12\catcode `\$12\catcode `\&12\catcode `\#12\catcode `\^12\catcode `\_12\catcode `\%12\relax}%
\providecommand \@@startlink[1]{}%
\providecommand \@@endlink[0]{}%
\providecommand \url  [0]{\begingroup\@sanitize@url \@url }%
\providecommand \@url [1]{\endgroup\@href {#1}{\urlprefix }}%
\providecommand \urlprefix  [0]{URL }%
\providecommand \Eprint [0]{\href }%
\providecommand \doibase [0]{https://doi.org/}%
\providecommand \selectlanguage [0]{\@gobble}%
\providecommand \bibinfo  [0]{\@secondoftwo}%
\providecommand \bibfield  [0]{\@secondoftwo}%
\providecommand \translation [1]{[#1]}%
\providecommand \BibitemOpen [0]{}%
\providecommand \bibitemStop [0]{}%
\providecommand \bibitemNoStop [0]{.\EOS\space}%
\providecommand \EOS [0]{\spacefactor3000\relax}%
\providecommand \BibitemShut  [1]{\csname bibitem#1\endcsname}%
\let\auto@bib@innerbib\@empty
\bibitem [{\citenamefont {Cowan}\ \emph {et~al.}(1956)\citenamefont {Cowan}, \citenamefont {Reines}, \citenamefont {Harrison}, \citenamefont {Kruse},\ and\ \citenamefont {McGuire}}]{Cowan:1956rrn}%
  \BibitemOpen
  \bibfield  {author} {\bibinfo {author} {\bibfnamefont {C.~L.}\ \bibnamefont {Cowan}}, \bibinfo {author} {\bibfnamefont {F.}~\bibnamefont {Reines}}, \bibinfo {author} {\bibfnamefont {F.~B.}\ \bibnamefont {Harrison}}, \bibinfo {author} {\bibfnamefont {H.~W.}\ \bibnamefont {Kruse}},\ and\ \bibinfo {author} {\bibfnamefont {A.~D.}\ \bibnamefont {McGuire}},\ }\href {https://doi.org/10.1126/science.124.3212.103} {\bibfield  {journal} {\bibinfo  {journal} {Science}\ }\textbf {\bibinfo {volume} {124}},\ \bibinfo {pages} {103} (\bibinfo {year} {1956})}\BibitemShut {NoStop}%
\bibitem [{\citenamefont {Vogel}\ and\ \citenamefont {Engel}(1989)}]{VogelEngel1989}%
  \BibitemOpen
  \bibfield  {author} {\bibinfo {author} {\bibfnamefont {P.}~\bibnamefont {Vogel}}\ and\ \bibinfo {author} {\bibfnamefont {J.}~\bibnamefont {Engel}},\ }\href {https://doi.org/10.1103/PhysRevD.39.3378} {\bibfield  {journal} {\bibinfo  {journal} {Phys. Rev. D}\ }\textbf {\bibinfo {volume} {39}},\ \bibinfo {pages} {3378} (\bibinfo {year} {1989})}\BibitemShut {NoStop}%
\bibitem [{\citenamefont {Beda}\ \emph {et~al.}(2012)\citenamefont {Beda} \emph {et~al.}}]{Beda2012}%
  \BibitemOpen
  \bibfield  {author} {\bibinfo {author} {\bibfnamefont {A.~G.}\ \bibnamefont {Beda}} \emph {et~al.},\ }\href {https://doi.org/10.1155/2012/350150} {\bibfield  {journal} {\bibinfo  {journal} {Adv. High Energy Phys.}\ }\textbf {\bibinfo {volume} {2012}},\ \bibinfo {pages} {350150} (\bibinfo {year} {2012})}\BibitemShut {NoStop}%
\bibitem [{\citenamefont {Abe}\ \emph {et~al.}(2022)\citenamefont {Abe} \emph {et~al.}}]{Super-Kamiokande:2021the}%
  \BibitemOpen
  \bibfield  {author} {\bibinfo {author} {\bibfnamefont {K.}~\bibnamefont {Abe}} \emph {et~al.} (\bibinfo {collaboration} {Super-Kamiokande Collaboration}),\ }\href {https://doi.org/10.1016/j.nima.2021.166248} {\bibfield  {journal} {\bibinfo  {journal} {Nucl. Instrum. Meth. A}\ }\textbf {\bibinfo {volume} {1027}},\ \bibinfo {pages} {166248} (\bibinfo {year} {2022})},\ \Eprint {https://arxiv.org/abs/2109.00360} {arXiv:2109.00360 [physics.ins-det]} \BibitemShut {NoStop}%
\bibitem [{\citenamefont {et~al.}(2003)}]{KamLAND2003}%
  \BibitemOpen
  \bibfield  {author} {\bibinfo {author} {\bibfnamefont {K.~E.}\ \bibnamefont {et~al.}} (\bibinfo {collaboration} {KamLAND Collaboration}),\ }\href {https://doi.org/10.1103/PhysRevLett.90.021802} {\bibfield  {journal} {\bibinfo  {journal} {Phys. Rev. Lett.}\ }\textbf {\bibinfo {volume} {90}},\ \bibinfo {pages} {021802} (\bibinfo {year} {2003})}\BibitemShut {NoStop}%
\bibitem [{\citenamefont {et~al.}(2005)}]{KamLAND2005}%
  \BibitemOpen
  \bibfield  {author} {\bibinfo {author} {\bibfnamefont {T.~A.}\ \bibnamefont {et~al.}} (\bibinfo {collaboration} {KamLAND Collaboration}),\ }\href {https://doi.org/10.1103/PhysRevLett.94.081801} {\bibfield  {journal} {\bibinfo  {journal} {Phys. Rev. Lett.}\ }\textbf {\bibinfo {volume} {94}},\ \bibinfo {pages} {081801} (\bibinfo {year} {2005})}\BibitemShut {NoStop}%
\bibitem [{\citenamefont {Dye}\ and\ \citenamefont {Barna}(2015)}]{geoneutrinospaperref}%
  \BibitemOpen
  \bibfield  {author} {\bibinfo {author} {\bibfnamefont {S.}~\bibnamefont {Dye}}\ and\ \bibinfo {author} {\bibfnamefont {A.}~\bibnamefont {Barna}},\ }\href@noop {} {\bibinfo {title} {Global antineutrino modeling for a web application}} (\bibinfo {year} {2015}),\ \Eprint {https://arxiv.org/abs/1510.05633} {arXiv:1510.05633 [physics.ins-det]} \BibitemShut {NoStop}%
\bibitem [{\citenamefont {Barna}\ and\ \citenamefont {Dye}()}]{geoneutrinos_accessed}%
  \BibitemOpen
  \bibfield  {author} {\bibinfo {author} {\bibfnamefont {A.~M.}\ \bibnamefont {Barna}}\ and\ \bibinfo {author} {\bibfnamefont {S.~T.}\ \bibnamefont {Dye}},\ }\href@noop {} {\bibinfo {title} {Antineutrino model}},\ \bibinfo {howpublished} {\url{https://reactors.geoneutrinos.org}},\ \bibinfo {note} {accessed 2 August 2025}\BibitemShut {NoStop}%
\bibitem [{\citenamefont {{International Atomic Energy Agency}}()}]{IAEA_PRIS}%
  \BibitemOpen
  \bibfield  {author} {\bibinfo {author} {\bibnamefont {{International Atomic Energy Agency}}},\ }\href@noop {} {\bibinfo {title} {{Power Reactor Information System (PRIS)}}},\ \bibinfo {howpublished} {\url{https://pris.iaea.org/pris/home.aspx}},\ \bibinfo {note} {accessed 2 August 2025}\BibitemShut {NoStop}%
\bibitem [{\citenamefont {Fukuda}\ \emph {et~al.}(2003)\citenamefont {Fukuda} \emph {et~al.}}]{Super-Kamiokande:2002weg}%
  \BibitemOpen
  \bibfield  {author} {\bibinfo {author} {\bibfnamefont {S.}~\bibnamefont {Fukuda}} \emph {et~al.} (\bibinfo {collaboration} {Super-Kamiokande Collaboration}),\ }\href {https://doi.org/10.1016/S0168-9002(03)00425-X} {\bibfield  {journal} {\bibinfo  {journal} {Nucl. Instrum. Meth. A}\ }\textbf {\bibinfo {volume} {501}},\ \bibinfo {pages} {418} (\bibinfo {year} {2003})}\BibitemShut {NoStop}%
\bibitem [{\citenamefont {Abe}\ \emph {et~al.}(2024{\natexlab{a}})\citenamefont {Abe} \emph {et~al.}}]{ABE2024169480}%
  \BibitemOpen
  \bibfield  {author} {\bibinfo {author} {\bibfnamefont {K.}~\bibnamefont {Abe}} \emph {et~al.} (\bibinfo {collaboration} {Super-Kamiokande Collaboration}),\ }\href {https://doi.org/10.1016/j.nima.2024.169480} {\bibfield  {journal} {\bibinfo  {journal} {Nucl. Instrum. Meth. A}\ }\textbf {\bibinfo {volume} {1065}},\ \bibinfo {pages} {169480} (\bibinfo {year} {2024}{\natexlab{a}})}\BibitemShut {NoStop}%
\bibitem [{\citenamefont {Beacom}\ and\ \citenamefont {Vagins}(2004)}]{PhysRevLett.93.171101}%
  \BibitemOpen
  \bibfield  {author} {\bibinfo {author} {\bibfnamefont {J.~F.}\ \bibnamefont {Beacom}}\ and\ \bibinfo {author} {\bibfnamefont {M.~R.}\ \bibnamefont {Vagins}},\ }\href {https://doi.org/10.1103/PhysRevLett.93.171101} {\bibfield  {journal} {\bibinfo  {journal} {Phys. Rev. Lett.}\ }\textbf {\bibinfo {volume} {93}},\ \bibinfo {pages} {171101} (\bibinfo {year} {2004})}\BibitemShut {NoStop}%
\bibitem [{\citenamefont {Carminati}(2015)}]{CARMINATI2015666}%
  \BibitemOpen
  \bibfield  {author} {\bibinfo {author} {\bibfnamefont {G.}~\bibnamefont {Carminati}},\ }\href {https://doi.org/10.1016/j.phpro.2014.12.068} {\bibfield  {journal} {\bibinfo  {journal} {Phys. Procedia}\ }\textbf {\bibinfo {volume} {61}},\ \bibinfo {pages} {666} (\bibinfo {year} {2015})}\BibitemShut {NoStop}%
\bibitem [{\citenamefont {Kato}\ \emph {et~al.}(2020)\citenamefont {Kato}, \citenamefont {Ishidoshiro},\ and\ \citenamefont {Yoshida}}]{Kato_2020}%
  \BibitemOpen
  \bibfield  {author} {\bibinfo {author} {\bibfnamefont {C.}~\bibnamefont {Kato}}, \bibinfo {author} {\bibfnamefont {K.}~\bibnamefont {Ishidoshiro}},\ and\ \bibinfo {author} {\bibfnamefont {T.}~\bibnamefont {Yoshida}},\ }\href {https://doi.org/10.1146/annurev-nucl-040620-021320} {\bibfield  {journal} {\bibinfo  {journal} {Annual Review of Nuclear and Particle Science}\ }\textbf {\bibinfo {volume} {70}},\ \bibinfo {pages} {121–145} (\bibinfo {year} {2020})}\BibitemShut {NoStop}%
\bibitem [{\citenamefont {Machado}\ \emph {et~al.}(2022)\citenamefont {Machado} \emph {et~al.}}]{Machado_2022}%
  \BibitemOpen
  \bibfield  {author} {\bibinfo {author} {\bibfnamefont {L.~N.}\ \bibnamefont {Machado}} \emph {et~al.} (\bibinfo {collaboration} {Super-Kamiokande Collaboration}),\ }\href {https://doi.org/10.3847/1538-4357/ac7f9c} {\bibfield  {journal} {\bibinfo  {journal} {Astrophys. J.}\ }\textbf {\bibinfo {volume} {935}},\ \bibinfo {pages} {40} (\bibinfo {year} {2022})},\ \Eprint {https://arxiv.org/abs/2205.09881} {arXiv:2205.09881 [hep-ex]} \BibitemShut {NoStop}%
\bibitem [{\citenamefont {Abe}\ \emph {et~al.}(2024{\natexlab{b}})\citenamefont {Abe} \emph {et~al.}}]{combined_preSN}%
  \BibitemOpen
  \bibfield  {author} {\bibinfo {author} {\bibfnamefont {S.}~\bibnamefont {Abe}} \emph {et~al.},\ }\href {https://doi.org/10.3847/1538-4357/ad5fee} {\bibfield  {journal} {\bibinfo  {journal} {Astrophys. J.}\ }\textbf {\bibinfo {volume} {973}},\ \bibinfo {pages} {140} (\bibinfo {year} {2024}{\natexlab{b}})}\BibitemShut {NoStop}%
\bibitem [{\citenamefont {Simpson}\ \emph {et~al.}(2019)\citenamefont {Simpson} \emph {et~al.}}]{charlespaper}%
  \BibitemOpen
  \bibfield  {author} {\bibinfo {author} {\bibfnamefont {C.}~\bibnamefont {Simpson}} \emph {et~al.} (\bibinfo {collaboration} {Super-Kamiokande Collaboration}),\ }\href {https://doi.org/10.3847/1538-4357/ab4883} {\bibfield  {journal} {\bibinfo  {journal} {Astrophys. J.}\ }\textbf {\bibinfo {volume} {885}},\ \bibinfo {pages} {133} (\bibinfo {year} {2019})}\BibitemShut {NoStop}%
\bibitem [{\citenamefont {Giunti}\ \emph {et~al.}(2022)\citenamefont {Giunti}, \citenamefont {Li}, \citenamefont {Ternes},\ and\ \citenamefont {Xin}}]{GIUNTI2022137054}%
  \BibitemOpen
  \bibfield  {author} {\bibinfo {author} {\bibfnamefont {C.}~\bibnamefont {Giunti}}, \bibinfo {author} {\bibfnamefont {Y.~F.}\ \bibnamefont {Li}}, \bibinfo {author} {\bibfnamefont {C.~A.}\ \bibnamefont {Ternes}},\ and\ \bibinfo {author} {\bibfnamefont {Z.}~\bibnamefont {Xin}},\ }\href {https://doi.org/10.1016/j.physletb.2022.137054} {\bibfield  {journal} {\bibinfo  {journal} {Phys. Lett. B}\ }\textbf {\bibinfo {volume} {829}},\ \bibinfo {pages} {137054} (\bibinfo {year} {2022})}\BibitemShut {NoStop}%
\bibitem [{\citenamefont {Kopeikin}\ \emph {et~al.}(2021)\citenamefont {Kopeikin}, \citenamefont {Skorokhvatov},\ and\ \citenamefont {Titov}}]{Kopeikin2021}%
  \BibitemOpen
  \bibfield  {author} {\bibinfo {author} {\bibfnamefont {V.}~\bibnamefont {Kopeikin}}, \bibinfo {author} {\bibfnamefont {M.}~\bibnamefont {Skorokhvatov}},\ and\ \bibinfo {author} {\bibfnamefont {O.}~\bibnamefont {Titov}},\ }\href {https://doi.org/10.1103/PhysRevD.104.L071301} {\bibfield  {journal} {\bibinfo  {journal} {Phys. Rev. D}\ }\textbf {\bibinfo {volume} {104}},\ \bibinfo {pages} {L071301} (\bibinfo {year} {2021})},\ \Eprint {https://arxiv.org/abs/2103.01684} {arXiv:2103.01684 [nucl-ex]} \BibitemShut {NoStop}%
\bibitem [{\citenamefont {Huber}(2011)}]{Huber2011}%
  \BibitemOpen
  \bibfield  {author} {\bibinfo {author} {\bibfnamefont {P.}~\bibnamefont {Huber}},\ }\href {https://doi.org/10.1103/PhysRevC.84.024617} {\bibfield  {journal} {\bibinfo  {journal} {Phys. Rev. C}\ }\textbf {\bibinfo {volume} {84}},\ \bibinfo {pages} {024617} (\bibinfo {year} {2011})},\ \bibinfo {note} {erratum: Phys. Rev. C 85, 029901 (2012)},\ \Eprint {https://arxiv.org/abs/1106.0687} {arXiv:1106.0687 [hep-ph]} \BibitemShut {NoStop}%
\bibitem [{\citenamefont {Estienne}\ \emph {et~al.}(2019)\citenamefont {Estienne} \emph {et~al.}}]{Estienne}%
  \BibitemOpen
  \bibfield  {author} {\bibinfo {author} {\bibfnamefont {M.}~\bibnamefont {Estienne}} \emph {et~al.},\ }\href {https://doi.org/10.1103/PhysRevLett.123.022502} {\bibfield  {journal} {\bibinfo  {journal} {Phys. Rev. Lett.}\ }\textbf {\bibinfo {volume} {123}},\ \bibinfo {pages} {022502} (\bibinfo {year} {2019})}\BibitemShut {NoStop}%
\bibitem [{\citenamefont {Mueller}\ \emph {et~al.}(2011)\citenamefont {Mueller} \emph {et~al.}}]{Mueller}%
  \BibitemOpen
  \bibfield  {author} {\bibinfo {author} {\bibfnamefont {T.~A.}\ \bibnamefont {Mueller}} \emph {et~al.},\ }\href {https://doi.org/10.1103/PhysRevC.83.054615} {\bibfield  {journal} {\bibinfo  {journal} {Phys. Rev. C}\ }\textbf {\bibinfo {volume} {83}},\ \bibinfo {pages} {054615} (\bibinfo {year} {2011})}\BibitemShut {NoStop}%
\bibitem [{\citenamefont {Strumia}\ and\ \citenamefont {Vissani}(2003)}]{StrumiaVissani2003}%
  \BibitemOpen
  \bibfield  {author} {\bibinfo {author} {\bibfnamefont {A.}~\bibnamefont {Strumia}}\ and\ \bibinfo {author} {\bibfnamefont {F.}~\bibnamefont {Vissani}},\ }\href {https://doi.org/10.1016/S0370-2693(03)00616-6} {\bibfield  {journal} {\bibinfo  {journal} {Phys. Lett. B}\ }\textbf {\bibinfo {volume} {564}},\ \bibinfo {pages} {42} (\bibinfo {year} {2003})},\ \Eprint {https://arxiv.org/abs/astro-ph/0302055} {arXiv:astro-ph/0302055} \BibitemShut {NoStop}%
\bibitem [{\citenamefont {{GLG4sim Collaboration}}(2006)}]{chooz_ref}%
  \BibitemOpen
  \bibfield  {author} {\bibinfo {author} {\bibnamefont {{GLG4sim Collaboration}}},\ }\href@noop {} {\bibinfo {title} {Additional gadolinium support for glg4sim}},\ \bibinfo {howpublished} {\url{https://www.phys.ksu.edu/personal/gahs/GLG4sim/Gd.html}} (\bibinfo {year} {2006}),\ \bibinfo {note} {accessed 19 June 2026}\BibitemShut {NoStop}%
\bibitem [{\citenamefont {Ou}\ \emph {et~al.}(2014)\citenamefont {Ou}, \citenamefont {Yano}, \citenamefont {Yamada}, \citenamefont {Mori}, \citenamefont {Kayano}, \citenamefont {Sakuda}, \citenamefont {Kimura},\ and\ \citenamefont {Harada}}]{anri_ref_sim}%
  \BibitemOpen
  \bibfield  {author} {\bibinfo {author} {\bibfnamefont {I.}~\bibnamefont {Ou}}, \bibinfo {author} {\bibfnamefont {T.}~\bibnamefont {Yano}}, \bibinfo {author} {\bibfnamefont {Y.}~\bibnamefont {Yamada}}, \bibinfo {author} {\bibfnamefont {T.}~\bibnamefont {Mori}}, \bibinfo {author} {\bibfnamefont {T.}~\bibnamefont {Kayano}}, \bibinfo {author} {\bibfnamefont {M.}~\bibnamefont {Sakuda}}, \bibinfo {author} {\bibfnamefont {A.}~\bibnamefont {Kimura}},\ and\ \bibinfo {author} {\bibfnamefont {H.}~\bibnamefont {Harada}},\ }\href {https://doi.org/10.7566/JPSCP.1.013053} {\bibfield  {journal} {\bibinfo  {journal} {JPS Conference Proceedings}\ }\textbf {\bibinfo {volume} {1}},\ \bibinfo {pages} {013053} (\bibinfo {year} {2014})}\BibitemShut {NoStop}%
\bibitem [{\citenamefont {Esteban}\ \emph {et~al.}(2024)\citenamefont {Esteban} \emph {et~al.}}]{NUFIT60}%
  \BibitemOpen
  \bibfield  {author} {\bibinfo {author} {\bibfnamefont {I.}~\bibnamefont {Esteban}} \emph {et~al.},\ }\href {https://doi.org/10.1007/JHEP12(2024)216} {\bibfield  {journal} {\bibinfo  {journal} {{J. High Energy Phys.}}\ }\textbf {\bibinfo {volume} {12}},\ \bibinfo {pages} {216} (\bibinfo {year} {2024})}\BibitemShut {NoStop}%
\bibitem [{\citenamefont {Abe}\ \emph {et~al.}(2024{\natexlab{c}})\citenamefont {Abe} \emph {et~al.}}]{PhysRevD.109.092001}%
  \BibitemOpen
  \bibfield  {author} {\bibinfo {author} {\bibfnamefont {K.}~\bibnamefont {Abe}} \emph {et~al.} (\bibinfo {collaboration} {Super-Kamiokande Collaboration}),\ }\href {https://doi.org/10.1103/PhysRevD.109.092001} {\bibfield  {journal} {\bibinfo  {journal} {Phys. Rev. D}\ }\textbf {\bibinfo {volume} {109}},\ \bibinfo {pages} {092001} (\bibinfo {year} {2024}{\natexlab{c}})}\BibitemShut {NoStop}%
\bibitem [{\citenamefont {{JUNO Collaboration}}(2026)}]{JUNO2026}%
  \BibitemOpen
  \bibfield  {author} {\bibinfo {author} {\bibnamefont {{JUNO Collaboration}}},\ }\href {https://doi.org/10.1038/s41586-026-10538-z} {\bibfield  {journal} {\bibinfo  {journal} {Nature}\ }\textbf {\bibinfo {volume} {654}},\ \bibinfo {pages} {343} (\bibinfo {year} {2026})}\BibitemShut {NoStop}%
\bibitem [{\citenamefont {et~al.}(2026)}]{SNOplus2026}%
  \BibitemOpen
  \bibfield  {author} {\bibinfo {author} {\bibfnamefont {M.~A.}\ \bibnamefont {et~al.}} (\bibinfo {collaboration} {SNO+ Collaboration}),\ }\href {https://doi.org/10.1103/mv5z-pfb7} {\bibfield  {journal} {\bibinfo  {journal} {Phys. Rev. Lett.}\ }\textbf {\bibinfo {volume} {137}},\ \bibinfo {pages} {101807} (\bibinfo {year} {2026})}\BibitemShut {NoStop}%
\end{thebibliography}%

\newpage

\onecolumngrid

\begin{center}
{\large\bfseries End Matter}
\end{center}

\twocolumngrid
\raggedbottom

\begin{figure}[h]
    \centering
    \includegraphics[width=0.75\linewidth]{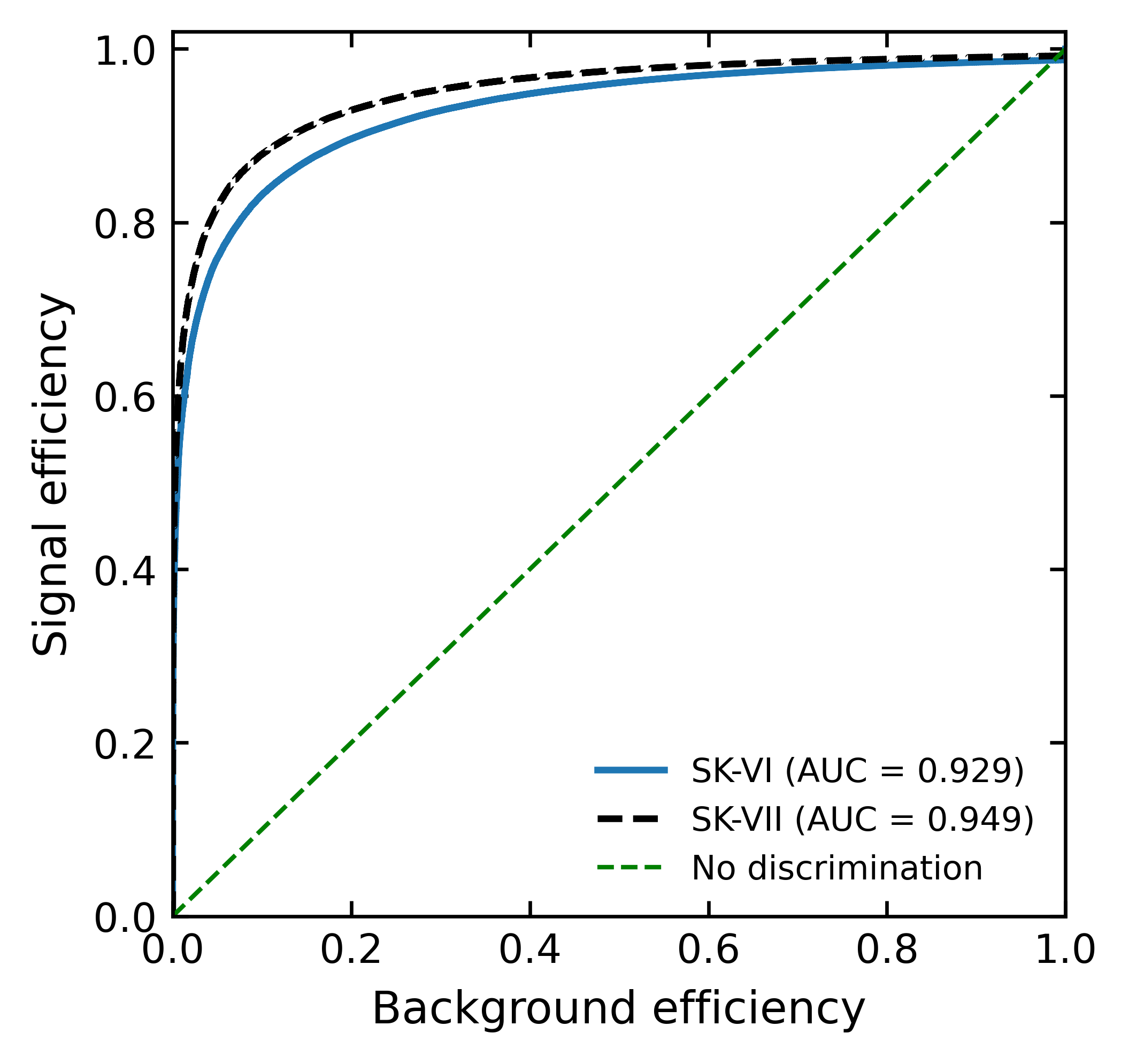}
    \label{fig:bdt_roc}
    \caption{
    ROC curves for the BDT selection in SK-VI and SK-VII. The area under
    the ROC curve (AUC) quantifies the classifier's ability to discriminate
    simulated IBD signal events from the background sample. The operating
    points used in the analysis are also shown.}
\end{figure}

\textit{Appendix A --- BDT event selection.}
The BDT cut is chosen by maximizing the figure of merit $S/\sqrt{S+B}$, where $S$ and $B$ are the expected signal and background yields after applying the cut. The optimization is performed independently for each SK-Gd phase. Figure~\ref{fig:bdt_roc} shows the corresponding receiver operating characteristic (ROC) curves for SK-VI and SK-VII.

\textit{Appendix B --- Reactor-antineutrino and geoneutrino predictions.}
The reactor-antineutrino and geoneutrino predictions used in this analysis were obtained from a local copy of~\citep{geoneutrinospaperref,geoneutrinos_accessed}. Rates are given in Neutrino Interaction Units (NIU), where $1~\mathrm{NIU}$ corresponds to one interaction per $10^{32}$ targets per year. Predictions evaluated using the NuFit 6.0~\citep{NUFIT60} oscillation parameters and the best-fit oscillation parameters obtained in this analysis ($\sin^2\theta_{12}=0.500$ and $\Delta m^2_{21}=9.08\times10^{-5}~\mathrm{eV}^2$) are shown. For the reactor prediction, the rates for SK-VI and SK-VII were obtained by averaging the monthly predictions over the months included in each analysis period. Both geoneutrino and reactor rates are evaluated over the prompt energy range $2<E_{e^+}<10~\mathrm{MeV}$, using $E_{e^+}=E_{\bar{\nu}_e}-1.293~\mathrm{MeV}$.

For the evaluated positron-energy range only the $^{238}\mathrm{U}$ decay chain contributes to the geoneutrino prediction, while the $^{232}\mathrm{Th}$ contribution is negligible. The integrated rates are $0.01083~\mathrm{NIU}$ using the NuFit 6.0 oscillation parameters and $0.00944~\mathrm{NIU}$ using the best-fit parameters obtained in this analysis. The corresponding binned geoneutrino spectra are shown in Table~\ref{tab:geoneutrino_spectrum}.

\begin{table}[h]
\centering
\caption{Binned geoneutrino interaction rate spectra for NuFit 6.0 and best-fit oscillation hypotheses used in this analysis. Rates are given in NIU/MeV. The predicted rate is zero for $E_{e^+}>4~\mathrm{MeV}$.}
\label{tab:geoneutrino_spectrum}
\begin{ruledtabular}
\begin{tabular}{
c
S[table-format=1.6]
S[table-format=1.6]
}
$E_{e^+}$ [MeV] & {NuFit 6.0} & {Best fit} \\
\hline
2--3 & 0.010558 & 0.009203 \\
3--4 & 0.000272 & 0.000237 \\
\end{tabular}
\end{ruledtabular}
\end{table}

The corresponding reactor interaction rate spectra are given in Table~\ref{tab:reactor_phase_spectra}. 

\begin{table}[h]
\centering
\caption{Reactor interaction rate spectra for SK-VI and SK-VII, obtained using the local copy of~\citep{geoneutrinospaperref,
geoneutrinos_accessed} and averaged over the months included in each analysis period. Rates are given in NIU/MeV.}
\label{tab:reactor_phase_spectra}
\begin{ruledtabular}
\begin{tabular}{
c
S[table-format=2.3]
S[table-format=2.3]
S[table-format=2.3]
S[table-format=2.3]
}
$E_{e^+}$ &
\multicolumn{2}{c}{SK-VI} &
\multicolumn{2}{c}{SK-VII} \\
\multicolumn{1}{c}{[MeV]} &
{NuFit 6.0} & {Best fit} &
{NuFit 6.0} & {Best fit} \\
\hline
2--3  & 22.097 & 28.615 & 35.045 & 37.378 \\
3--4  & 28.155 & 13.046 & 42.927 & 22.155 \\
4--5  & 19.400 & 14.611 & 27.367 & 22.793 \\
5--6  & 7.040  & 8.769  & 9.533  & 12.394 \\
6--7  & 1.491  & 1.675  & 1.910  & 2.334  \\
7--8  & 0.095  & 0.175  & 0.126  & 0.220  \\
8--9  & 0.016  & 0.020  & 0.021  & 0.025  \\
9--10 & 0.000  & 0.000  & 0.000  & 0.000  \\
\end{tabular}
\end{ruledtabular}
\end{table}

Tables~\ref{tab:monthly_reactor_skvi} and \ref{tab:monthly_reactor_skvii} give the monthly reactor interaction rates for SK-VI and SK-VII, respectively. The SK-VI table also includes the predictions for November 2020--January 2021 for reference: November and December 2020 correspond to the low-reactor period used for the background estimate, while January 2021 is excluded from the analysis because nearby reactors were ramping up.

\begin{table}[h]
\centering
\caption{Monthly reactor interaction rates for SK-VI, obtained using the local copy of~\citep{geoneutrinospaperref, geoneutrinos_accessed} and integrated over $2<E_{e^+}<10~\mathrm{MeV}$. Rates are given in NIU.}
\label{tab:monthly_reactor_skvi}
\begin{ruledtabular}
\begin{tabular}{
l
c
S[table-format=3.2]
S[table-format=3.2]
}
Month & Year & {NuFit 6.0} & {Best fit} \\
      &      & {[NIU]}     & {[NIU]} \\
\hline
November  & 2020 & 28.20  & 24.59 \\
December  & 2020 & 28.82  & 25.24 \\
January   & 2021 & 36.64  & 31.56 \\
February  & 2021 & 47.73  & 40.54 \\
March     & 2021 & 57.83  & 49.75 \\
April     & 2021 & 68.50  & 59.63 \\
May       & 2021 & 75.17  & 65.85 \\
June      & 2021 & 74.01  & 64.78 \\
July      & 2021 & 117.34 & 97.80 \\
August    & 2021 & 121.08 & 100.74 \\
September & 2021 & 122.40 & 101.72 \\
October   & 2021 & 115.48 & 96.74 \\
November  & 2021 & 95.89  & 82.46 \\
December  & 2021 & 101.37 & 87.40 \\
January   & 2022 & 101.57 & 87.61 \\
February  & 2022 & 99.52  & 85.78 \\
March     & 2022 & 69.67  & 59.92 \\
April     & 2022 & 62.33  & 53.92 \\
May       & 2022 & 62.56  & 54.08 \\
\end{tabular}
\end{ruledtabular}
\end{table}

\begin{table}[h]
\centering
\caption{Monthly reactor interaction rates for SK-VII, obtained using the local copy of~\citep{geoneutrinospaperref, geoneutrinos_accessed} and integrated over $2<E_{e^+}<10~\mathrm{MeV}$. Rates are given in NIU.}
\label{tab:monthly_reactor_skvii}

\renewcommand{\arraystretch}{0.92}
\begin{ruledtabular}
\begin{tabular}{
l
c
S[table-format=3.2]
S[table-format=3.2]
}
Month & Year & {NuFit 6.0} & {Best fit} \\
      &      & {[NIU]}     & {[NIU]} \\
\hline
September & 2022 & 87.86  & 72.53 \\
October   & 2022 & 91.22  & 75.10 \\
November  & 2022 & 102.55 & 85.51 \\
December  & 2022 & 115.21 & 96.24 \\
January   & 2023 & 127.27 & 105.98 \\
February  & 2023 & 114.03 & 93.81 \\
March     & 2023 & 115.18 & 94.83 \\
April     & 2023 & 124.44 & 103.38 \\
May       & 2023 & 125.60 & 104.35 \\
June      & 2023 & 126.67 & 105.32 \\
July      & 2023 & 127.42 & 106.07 \\
August    & 2023 & 138.80 & 116.66 \\
September & 2023 & 122.88 & 104.25 \\
\end{tabular}
\end{ruledtabular}
\end{table}

\textit{Appendix C --- Statistical treatment.}
The oscillation parameters are determined by comparing the observed prompt-energy spectra with predictions over the
$(\sin^2\theta_{12},\,\Delta m^2_{21})$ parameter space and minimizing a~$\chi^2$ function including statistical and systematic uncertainties.
The fit therefore varies two oscillation parameters, $\sin^2\theta_{12}$ and $\Delta m^2_{21}$, while
$\sin^2\theta_{13}$ is fixed to the NuFit v6.0 best-fit value.

For the combined SK-VI and SK-VII spectral goodness-of-fit, correlations between the two data-taking periods are included using Gaussian-constrained nuisance parameters. The corresponding $\chi^2$ is written as
\begin{equation}
\chi^2(\boldsymbol{\xi}) =
\sum_{p,i}
\frac{
\left[
O_{pi}-P_{pi}-\sum_k \xi_k\Delta_{pi}^{(k)}
\right]^2
}
{\sigma_{pi,\mathrm{stat}}^2}
+
\sum_k \xi_k^2 ,
\label{eq:chi2_pull}
\end{equation}
where $p$ denotes the SK-Gd phase, $i$ the reconstructed prompt-energy bin, $O_{pi}$ and $P_{pi}$ are the observed and predicted event rates, respectively, $\sigma_{pi,\mathrm{stat}}$ is the statistical uncertainty of the observed event count, and $\xi_k$ are nuisance parameters describing uncertainties in the prediction. $\Delta_{pi}^{(k)}$ represents one-standard-deviation change associated with uncertainty $k$. 

Common model uncertainties are described by nuisance parameters shared between SK-VI and SK-VII, while detector-related uncertainties are treated separately for the two phases. The reactor-antineutrino prediction, IBD cross section, neutron-capture $\gamma$-ray modeling, and geoneutrino prediction uncertainties are treated as correlated between SK-VI and SK-VII. Trigger efficiency, live time and event-selection efficiency are treated independently between the two phases. The systematic uncertainties included in the spectral analysis are summarized in Table~\ref{tab:systematics}.

\begin{table}[t]
\caption{\label{tab:systematics}
Summary of the systematic uncertainties included in the reactor antineutrino analysis.}
\begin{ruledtabular}
\begin{tabular}{lc}
Source & Uncertainty \\
\hline
Reactor antineutrino prediction & 3.0\% \\
IBD cross section & 0.4\% \\
$\gamma$-ray modeling $\gamma$-ray modeling & 5.4\% \\
Geoneutrino prediction & 25\% \\
Trigger efficiency & 3.0--7.0\% \\
IBD coincidence-search efficiency & 5.5\% \\
Other detector/selection uncertainties & 1.15\% \\
\end{tabular}
\end{ruledtabular}
\end{table}

The accidental-background estimate has an energy-bin-dependent uncertainty dominated by the statistical uncertainty of the low-reactor data sample. Since the estimates for SK-VI and SK-VII are obtained from the same low-reactor dataset, the uncertainties in corresponding energy bins are conservatively treated as correlated between the two phases.

The goodness-of-fit uses eight reconstructed prompt-energy bins from each SK-Gd phase, giving 16 bins in total. The prediction is evaluated at the best-fit oscillation point, while the nuisance parameters are constrained by their Gaussian pull terms and are not treated as unconstrained fitted parameters in the quoted number of degrees of freedom. The goodness-of-fit is therefore evaluated with 16 degrees of freedom, giving $\chi^2=16.7$ and a~$p$-value of 0.40. As a cross-check, the combined spectral goodness-of-fit was also evaluated using a~covariance-matrix formulation containing the same uncertainties and correlations. The Gaussian-constrained nuisance and covariance-matrix formulations give consistent results.

The reactor-antineutrino signal significance is evaluated from
\begin{equation}
\Delta\chi^2_{\rm reac}
=
\chi^2_{\rm no\ reactor}
-
\chi^2_{\rm osc}.
\end{equation}
The no-reactor hypothesis gives $\Delta\chi^2_{\rm reac}=34.7$ relative to the oscillated reactor-antineutrino prediction, corresponding to an approximate significance of approximately $6\sigma$.

\textit{Appendix D --- Solar and reactor consistency.}
The consistency between the reactor and solar neutrino measurements is evaluated using a~parameter goodness-of-fit test. The test statistic is defined as
\begin{equation}
\Delta\chi^2_{\rm PG}
=
\chi^2_{\rm comb,min}
-
\chi^2_{\rm reac,min}
-
\chi^2_{\rm solar,min},
\end{equation}
where $\chi^2_{\rm comb,min}$ is the minimum of the combined solar and reactor fit, and $\chi^2_{\rm reac,min}$ and $\chi^2_{\rm solar,min}$ are the minima obtained from the individual datasets. Since the two datasets share the parameters $\sin^2\theta_{12}$ and $\Delta m^2_{21}$, the PG statistic is evaluated with two degrees of freedom. The resulting $\Delta\chi^2_{\rm PG}=7.64$ corresponds to a~$p$-value of 0.022, implying Gaussian-equivalent compatibility between the two datasets of approximately 1.9$\sigma$.

\end{document}